\documentclass[aps,amsmath,amssymbols,superscriptaddress,11pt]{revtex4}
\usepackage[]{graphicx}
\usepackage{color}
\usepackage{bbold}
\usepackage{braket}
\usepackage[colorlinks=true, pdfstartview=FitV, linkcolor=blue, citecolor=red, urlcolor=black]{hyperref}
\usepackage{orcidlink}
\newcommand{\be}{\begin{equation}}
\newcommand{\ee}{\end{equation}}
\newcommand{\ben}{\begin{eqnarray}}
\newcommand{\een}{\end{eqnarray}}
\newcommand{\bes}{\begin{subequations}}
\newcommand{\ees}{\end{subequations}}
\def\bal#1\eal{\begin{align}#1\end{align}}
\usepackage{comment}

\newcommand{\bfi}{\begin{figure}}
\newcommand{\efi}{\end{figure}}
\newcommand{\bc}{\begin{center}}
\newcommand{\ec}{\end{center}}
\newcommand{\sech}{\mbox{sech}}

\newcommand{\arcsinh}{\mbox{arcsinh}}

\begin{document}
    \title{Geometrically constrained kinklike configurations engendering\\ long range, double exponential, half-compact and compact behavior}
    \author{D. Bazeia\,\orcidlink{0000-0003-1335-3705}}
     \email[]{bazeia@fisica.ufpb.br}
    \affiliation{Departamento de F\'\i sica, Universidade Federal da Para\'\i ba, 58051-970 Jo\~ao Pessoa, PB, Brazil}
    
        \author{M.A. Marques\,\orcidlink{0000-0001-7022-5502}}
        \email[]{marques@cbiotec.ufpb.br}
    \affiliation{Departamento de Biotecnologia, Universidade Federal da Para\'iba, 58051-900 Jo\~ao Pessoa, PB, Brazil}
    
    \author{R. Menezes\,\orcidlink{0000-0002-9586-4308}}
     \email[]{rmenezes@dcx.ufpb.br}
    \affiliation{Departamento de Ci\^encias Exatas, Universidade Federal da Para\'iba, 58297-000 Rio Tinto, PB, Brazil}

\begin{abstract}
We describe a procedure that contributes to modify the asymptotic behavior of kinks in a model described by two real scalar fields. The investigation takes advantage of a first order formalism based on energy minimization to unveil how to modify the asymptotic profile of kinklike configurations. In particular, we show that the exponential tails of standard kinklike configurations can be  smoothly modified to engender long range, double exponential, half-compact or compact behavior.
\end{abstract}
\maketitle

\section{Introduction} 

In high energy physics, models described by real scalar fields are of current interest since they may find distinct applications of practical use, as reported, for instance, in Refs. \cite{B1,B2,B3}. In $1+1$ spacetime dimensions, scalar fields are known to allow for the presence of localized spatial structures that attain topological profile. They are generically known as kinks, and an important prototype is the model
\be
\mathcal{L}_s = \frac12\partial_\mu\phi\partial^\mu\phi -\frac12(1-\phi^2)^2.
\ee
Here, the field $\phi=\phi(x,t)$ and the time and the spatial coordinates are all dimensionless.

This model describes static and uniform field configurations at $\bar{\phi}_{\pm}=\pm1$, which are minima of the potential. Also, there are nontrivial static configurations that represent kink $(+)$ and antikink $(-)$, exactly described by $ 
\phi_{\pm}=\pm \tanh(x)
$. Since they have the same asymptotic profile, we shall consider
\be 
\phi(x)=\tanh(x)
\ee 
as the kink configuration of the above model. Furthermore, one notices that the solution connects the minima of the potential, and has exponential tails that obey  $\phi_{asy}\mp1\propto \exp(-m_\pm |x|)$, where $m^2_\pm = V_{\phi\phi}(\pm1)$ and $\phi_{asy}$ stands for the asymptotic behavior of the kink. The above potential engenders spontaneous symmetry breaking and is known to represent the $\phi^4$ model. It has been investigated before in Refs.~\cite{B1,B2,B3} and more recently in \cite{B4} with many distinct motivations. It may also be generalized and used in several other circumstances, in particular to describe solitons in conducting polymers \cite{CP}, in supersymmetric systems \cite{SUSY}, and in Bose-Einstein condensates \cite{BE,nature} as dark and bright solitons. In magnetic materials, it can be considered to describe domain walls separating regions of distinct magnetic domains, as reported before in Ref. \cite{B5}.

Turning attention to the profile of the localized structure, along the years several investigations have dealt with the asymptotic behavior of kinks \cite{arodz,quasi,gaeta,highly,kinktocomp,comp,hybrid,hybrid2,highly2,gani,az,Khare,Khare0}, which is directly related to the second derivative of the potential at its minima. In particular, in Refs.~\cite{highly,highly2} the authors studied models in which $m_\pm=0$, supporting kinks with power-law tails. The slow falloff of these structures may give rise to long range interactions. One may also find the possibility of having double exponential tails or compact solutions, which appear for potentials with narrow minima, such that $m_\pm\to\infty$; see, e.g., Refs.~\cite{B4,kinktocomp,comp,hybrid} and references therein.

The interest in such localized structures is diverse; for instance, the dipole-dipole interaction is power-law or long range, and with Rydberg atoms one may build atomic structures with large electric dipole moments, so highlighting the long range characteristics of the interaction, which finds distinct application of practical use, in particular in atomic, molecular and optical physics; see, e.g., Ref. \cite{Book} and references therein. Also, long range, quasi-compact and compact structures are among the building blocks for modelling of colloidal self-assembly; see, e.g., Ref. \cite{Nature} for a recent review article concerned with the design of soft materials such as colloids and nanoparticles, in another line of research of current interest. We can also consider the new structures to be described in this work to study collisions, to understand how the modifications may induce different behavior in the scattering, modifying the fractal structure of the collisions in a way similar to the cases recently investigated in Refs. \cite{gani,sca1,S1,S2,S3}.

\section{General considerations}

In order to simulate the effects of  geometric constrictions that can be introduced in magnetic materials \cite{jubert}, in Refs. \cite{multikink,multikink2} the inclusion of a second scalar field was considered. In this case, the model is described by the Lagrangian density
\be\label{lagr}
\mathcal{L} = \frac12f(\chi)\partial_\mu\phi\partial^\mu\phi + \frac12\partial_\mu\chi\partial^\mu\chi - V(\phi,\chi).
\ee
The presence of a non negative $f(\chi)$ allows for modifications in the derivative at some points of the kink, making it vanish. This can lead to a solution with interesting internal structure, so the $\chi$ field may be used with the function $f(\chi)$ to modify the core of the structure. Since this appears inside the kinklike configuration, here we are thinking of bringing another motivation, dealing with the tail of the kink, which was not investigated before.

In this sense, in the present work we investigate a new class of functions $f(\chi)$ that is appropriate to modify the exponential tails of the standard kink associated to the $\phi^4$ model. We shall follow the lines of Refs.~\cite{bogo,multikink}, so we take the first order equations that minimize the energy of the system. The model \eqref{lagr} gives rise to the equations of motion 
\bes\bal
&\partial_\mu(f(\chi)\partial^\mu\phi)+V_\phi=0,
\\
&\partial_\mu\partial^\mu\chi-\frac{1}{2}\frac{df}{d\chi}\partial_\mu\phi\partial^\mu\phi + V_\chi=0,
\eal
\ees
where $V_\phi=\partial V/\partial \phi$ and $V_\chi=\partial V/\partial\chi $.
In the case of static fields, they become
\begin{subequations}\label{eoms}
		\begin{align}
			\label{est1} &f(\chi)\;\frac{d^2\phi}{dx^2}+\frac{df}{d\chi}\,\frac{d\phi}{dx}\,\frac{d\chi}{dx}=V_{\phi}, \\ 
			\label{est2}&\frac{d^2\chi}{dx^2} - \frac{1}{2}\frac{df}{d\chi}\,\left(\frac{d\phi}{dx} \right)^ 2 = V_{\chi}.
		\end{align}
	\end{subequations}
Also, the energy density associated to the static fields is
\begin{equation}
 \rho=\frac{f(\chi)}{2}\left(\frac{d\phi}{dx}\right)^2 +\frac12 \left(\frac{d\chi}{dx}\right)^2 + V(\phi,\chi). \label{T00} \\
\end{equation}
We can introduce an auxiliary function $W=W(\phi,\chi)$ which help us rewrite this energy density as
\begin{equation}
\rho =\frac{f(\chi)}{2}\left(\frac{d\phi}{dx} \mp \frac{W_{\phi}}{f(\chi)}\right)^2 + \frac{1}{2}\left(\frac{d\chi}{dx} \mp W_{\chi}\right)^2 + V-\left(\frac{1}{2}\frac{W_{\phi}^2}{f(\chi)}+\frac{1}{2}W_{\chi}^2\right) \pm\frac{dW}{dx},
\end{equation}
where $W_\phi=\partial W/\partial\phi$ and $W_\chi=\partial W/\partial \chi$. 
If the potential has the form
\be
V(\phi,\chi)=\frac12\frac{W_{\phi}^2}{f(\chi)} + \frac12\,{W_{\chi}^2},
\ee
the model can be described by the first order equations $d\phi/dx=\pm W_{\phi}/f(\chi)$ and $d\chi/dx=\pm W_{\chi}$. Solutions that obey these expressions minimize the energy of the system to $E = \left|W(v_+,w_+)-W(v_-,w_-)\right|$, where $v_\pm = \phi(\pm\infty)$ and $w_\pm=\chi(\pm\infty)$. We then consider the special case in which $W$ is separable in the form $W(\phi,\chi)=W(\phi)+g(\chi).$
In this situation, one can introduce the new coordinate $\xi=\xi(x)$ such that
\be\label{xidef}
\frac{d\xi}{dx}= \frac{1}{f(\chi(x))}.
\ee
It allows to write the first order equations as
\be\label{fowphichixi}
\frac{d\phi}{d\xi}=\pm W_{\phi}(\phi),\qquad
			\frac{d\chi}{dx}=\pm g_{\chi}(\chi),
\ee
where $g_\chi=dg/d\chi$. The energy density can be separated in two terms, in the form $\rho = \rho_1 + \rho_2$, where
\be\label{rho12}
\rho_1 = f(\chi)\left(\frac{d\phi}{dx}\right)^2,\quad \rho_2 = \left(\frac{d\chi}{dx}\right)^2.
\ee
To investigate how $\chi$ modifies the asymptotic behavior of the $\phi^4$ model via the function $f(\chi)$, we take the auxiliary function as
\be\label{W}
 W(\phi,\chi) = \phi-\frac13\phi^3 + g(\chi),
\ee
with $g(\chi)$ being a function that drives the first order equation for $\chi$; see Eq.~\eqref{fowphichixi}. For the above choice, $\phi$ is governed by $d\phi/d\xi = \pm (1-\phi^2)$. Since the equations with upper/lower sign are related by the change $x\to-x$, we take the one with positive sign, which supports the solution
\be\label{tanhxi}
\phi(x) = \tanh(\xi),
\ee
where $\xi$ is as in Eq.~\eqref{xidef} for a given function $f(\chi)$. For $\xi=x$, one gets that  $f=1$, and this recovers the standard $\phi^4$ kink, as expected. To calculate the coordinate $\xi$, one has to provide the functions $f(\chi)$ and $g(\chi)$. By integrating the right equation in \eqref{fowphichixi}, we get 
\be\label{fochig}
\pm (x-x_0)=\int \frac{d\chi}{g_\chi}=F(\chi)\; \Rightarrow\; \chi(x) = F^{-1}(x-x_0).
\ee
The functions $f$ and $g$ must lead to a coordinate $\xi(x)$ that ranges from $-\infty$ to $+\infty$, in order to keep the $\phi$ solution connecting the minima of the potential. In Ref.~\cite{multikink}, where the interest was to modify exclusively the core of the structure, one always has $f(w_\pm)=1$, which leads to $\xi \approx \xi_0 + kx$ for large values of $x$, where $\xi_0$ and $k$ are real constants, preserving the asymptotic behavior of the standard solution. In the present work, we shall take the functions $f$ and $g$ to explore other features of the model, modifying the asymptotic behavior of the solutions.

\section{Models} 

Let us now investigate some specific models, searching for solutions that engender distinct asymptotic behavior.

\subsection{Solution with power-law tail}

\begin{figure}[t!]
	\includegraphics[width=6.2cm]{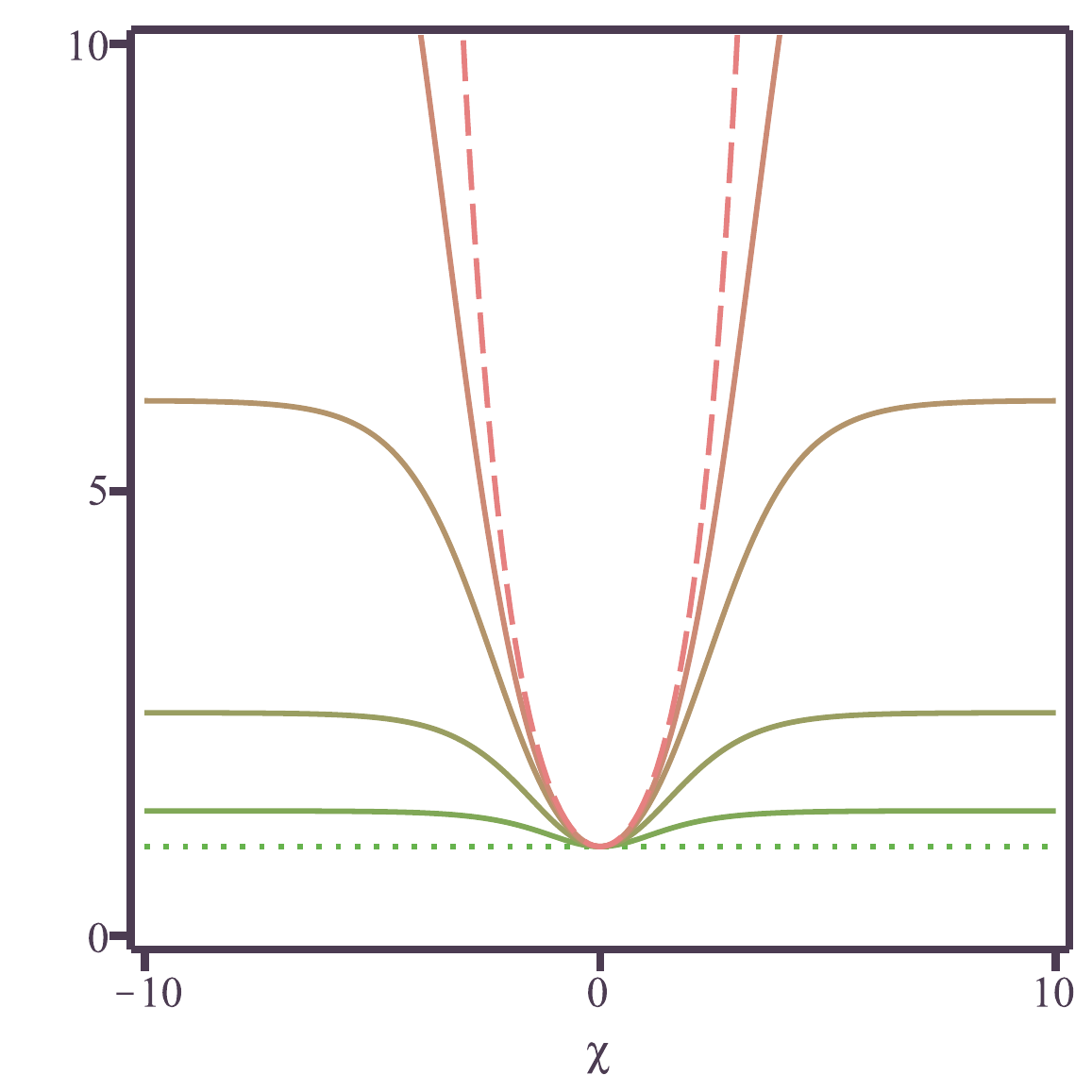}
	\includegraphics[width=6.2cm]{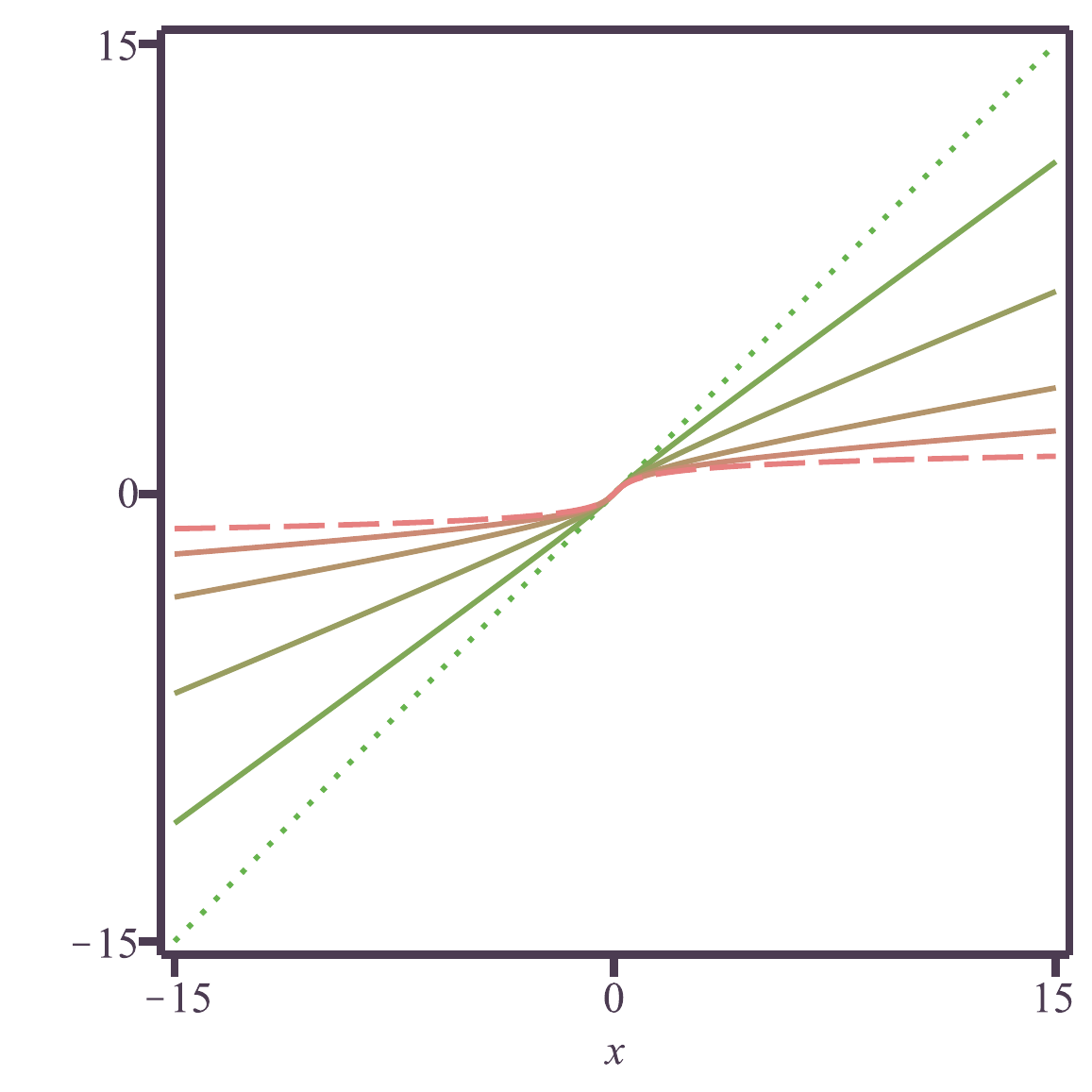}\\
	\includegraphics[width=6.2cm]{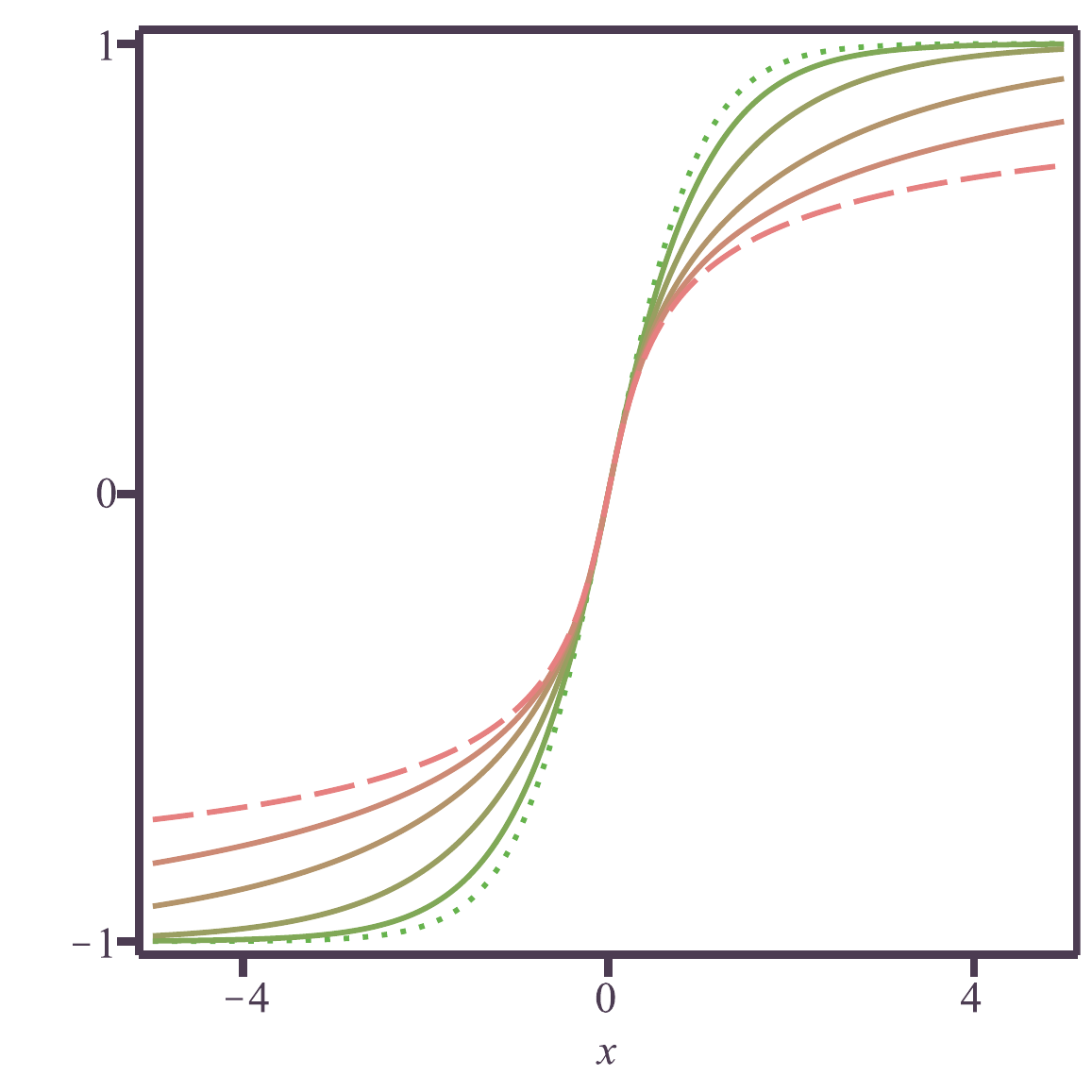}
	\includegraphics[width=6.2cm]{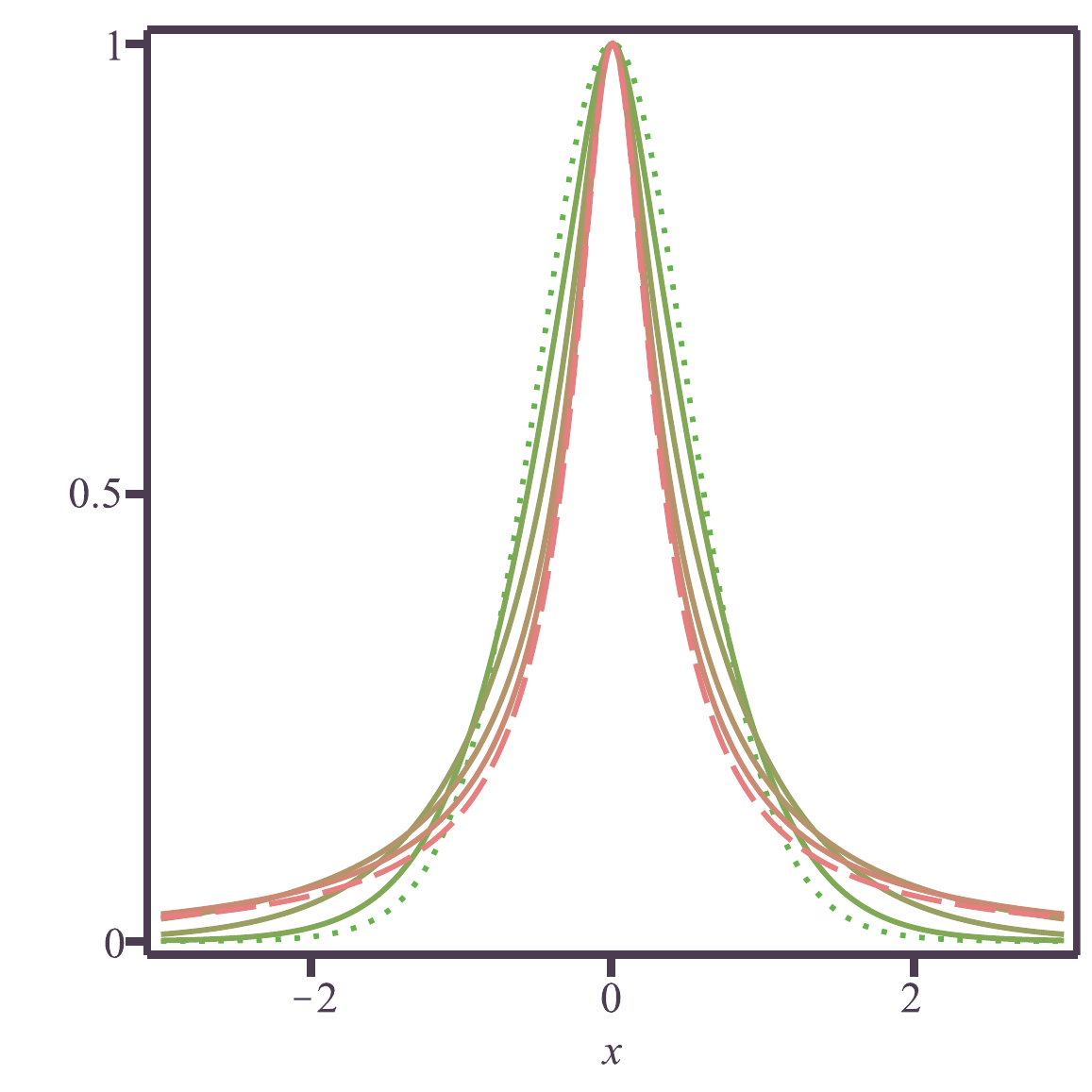}
	\caption{The function $f(\chi)$ (top left), the coordinate $\xi$ (top right), the solution $\phi$ (bottom left) and the energy density $\rho_1$ (bottom right) associated to the model described by Eqs.~\eqref{f1} and \eqref{gvac} for $\alpha=4$, $\xi_0=0$ and for $\beta=0, 0.4, 1.5, 5, 15$, and $\beta\to\infty$. The dotted lines represent the case $\beta=0$ and the dashed ones stand for $\beta\to\infty$.}
	\label{firstmodel}
\end{figure}

We first introduce
\be\label{f1}
f(\chi) = \frac{1+\beta}{1+\beta\,\sech(\chi)},
\ee
with $\beta$ being a real and non negative parameter. One then sees that $f(\chi)\geq1$ and $f(0)=1$. In particular, for $\beta=0$ we have $f=1$, where the fields are decoupled, and for $\beta\to\infty$ we have $f(\chi)=\cosh(\chi)$. At the boundary values of $\chi$, one has $f(w_\pm) = (1+\beta)/(1+\beta\sech(w_\pm))$. This expression shows us the following feature: $w_\pm$ controls how large $f$ gets asymptotically. This means that, for $w_\pm\to\infty$, $f$ becomes unbounded in the $\beta\to\infty$ case. In Fig. \ref{firstmodel} we depict the main results for this model, described by 
the functions $f(\chi)$ and $g(\chi)$ in Eqs. \eqref{f1} and \eqref{gvac}, respectively. We display the function $f(\chi)$, the coordinate $\xi$, the solution $\phi$ and the energy density $\rho_1$ associated to the model described by Eqs.~\eqref{f1} and \eqref{gvac}. When necessary, we take $\alpha=4$, $\xi_0=0$ and $\beta=0, 0.4, 1.5, 5, 15$, and $\beta\to\infty$.
In particular, in the top left panel of Fig.~\ref{firstmodel}, we display the above function $f(\chi)$. A model that support the aforementioned behavior for its solutions is the so-called vacuumless model, described by
\be\label{gvac}
g(\chi)= \alpha\arctan(\sinh(\chi)).
\ee
It was studied previously in Refs.~\cite{Vi,vacuumless}. Using Eq.~\eqref{fochig}, one gets
\be\label{solchi1}
\chi(x) = \arcsinh(\alpha x).
\ee
Notice that $\chi(\pm\infty) \to \pm\infty$, as we wanted to have an unbounded $f$. The energy density $\rho_2$ associated to this solution can be calculated from Eq.~\eqref{rho12}, which reads
\be
\rho_2(x) = \frac{\alpha^2}{1+\alpha^2x^2}.
\ee

The solution \eqref{solchi1}, combined with the function \eqref{f1}, must be used in Eq.~\eqref{xidef} to determine the coordinate $\xi$ of the hyperbolic tangent in Eq.~\eqref{tanhxi}. It has the form
\be\label{xi1}
\xi = \frac{1}{\beta+1}\left(x+\frac{\beta}{\alpha}\arcsinh(\alpha x) \right) + \xi_0.
\ee
From this expression, we see that the term in $x$ dominates at infinity for a general $\beta$. However, as $\beta$ gets larger and larger, this term becomes less and less relevant compared to the $\arcsinh(\alpha x)$ term such that, in the limit $\beta\to\infty$, it vanishes and we are left with
\be\label{xi1inf}
\xi = \frac1\alpha\,\arcsinh(\alpha x) + \xi_0,
\ee
which engender a logarithmic asymptotic behavior. In the top right panel of Fig.~\ref{firstmodel}, one can see the behavior of $\xi$ for several values of $\beta$, showing a transition from the linear ($\beta=0$) to the logarithmic type ($\beta\to\infty$). 

The solution is calculated combining Eqs.~\eqref{tanhxi} and \eqref{xi1}. The symmetric solution ($\xi_0=0$) can be seen from the curves at the bottom left panel of Fig.~\ref{firstmodel}. Notice that, near the origin, the behavior is the same, $\phi(x)\propto x$, for all $\beta$, so the modifications appear exclusively in the tail. For the $\beta\to\infty$ limit, since $\xi$ is given by \eqref{xi1inf}, the solution takes the form
\be
\phi(x) = \frac{\left(\alpha x + \sqrt{1+\alpha^2x^2}\right)^{2/\alpha} - C}{\left(\alpha x + \sqrt{1+\alpha^2x^2}\right)^{2/\alpha} + C},
\ee
where we have defined $C=\exp(-2\xi_0)$. The above expression shows that this solution engender a long range tail, depicted for $C=1$ in the dashed line of the bottom left panel at Fig.~\ref{firstmodel}, which may lead to highly interactive configurations, as described in Ref.~\cite{highly}. Indeed, one can expand the above expression far away from the origin, $x\to\pm\infty$, to show that the solution has power-law tails, as
\be
\phi_\pm(x) \approx \pm1 \mp\frac{2\delta_\pm}{|x|^{2/\alpha}}\pm\frac{2\delta_\pm^2}{|x|^{4/\alpha}} + \mathcal{O}\left[\frac{1}{|x|^{6/\alpha}}\right],
\ee
where $\delta_+ = C/(2\alpha)^{2/\alpha}$ and $\delta_- = 1/(C (2\alpha)^{2/\alpha})$. For $C=1$, we have $\delta_+=\delta_-$, so the tails are symmetric.
The contribution $\rho_1$ in the energy density \eqref{rho12} is
\be
\rho_1(x) = \frac{1}{1+\beta}\left(1+\frac{\beta}{\sqrt{1+\alpha^2x^2}}\right)\sech^4(\xi).
\ee
In the limit $\beta\to\infty$, the above expression takes the form
\be
\rho_1(x) = \frac{16C^2}{\left(1+\alpha^2x^2\right)^{3/2}} \frac{\left(\alpha x + \sqrt{1+\alpha^2x^2}\right)^{4/\alpha}}{\left(\left(\alpha x + \sqrt{1+\alpha^2x^2}\right)^{2/\alpha} + C \right)^4},
\ee
with power-law tails. The energy density can be seen in the bottom right panel of Fig.~\ref{firstmodel} for several values of $\beta$.
The functions in Eqs.~\eqref{f1} and \eqref{gvac} smoothly makes the falloff of the solution become slower, going from exponential to power-law tails. 

\subsection{Solutions with double exponential tail}

\begin{figure}[t!]
	\includegraphics[width=6.2cm]{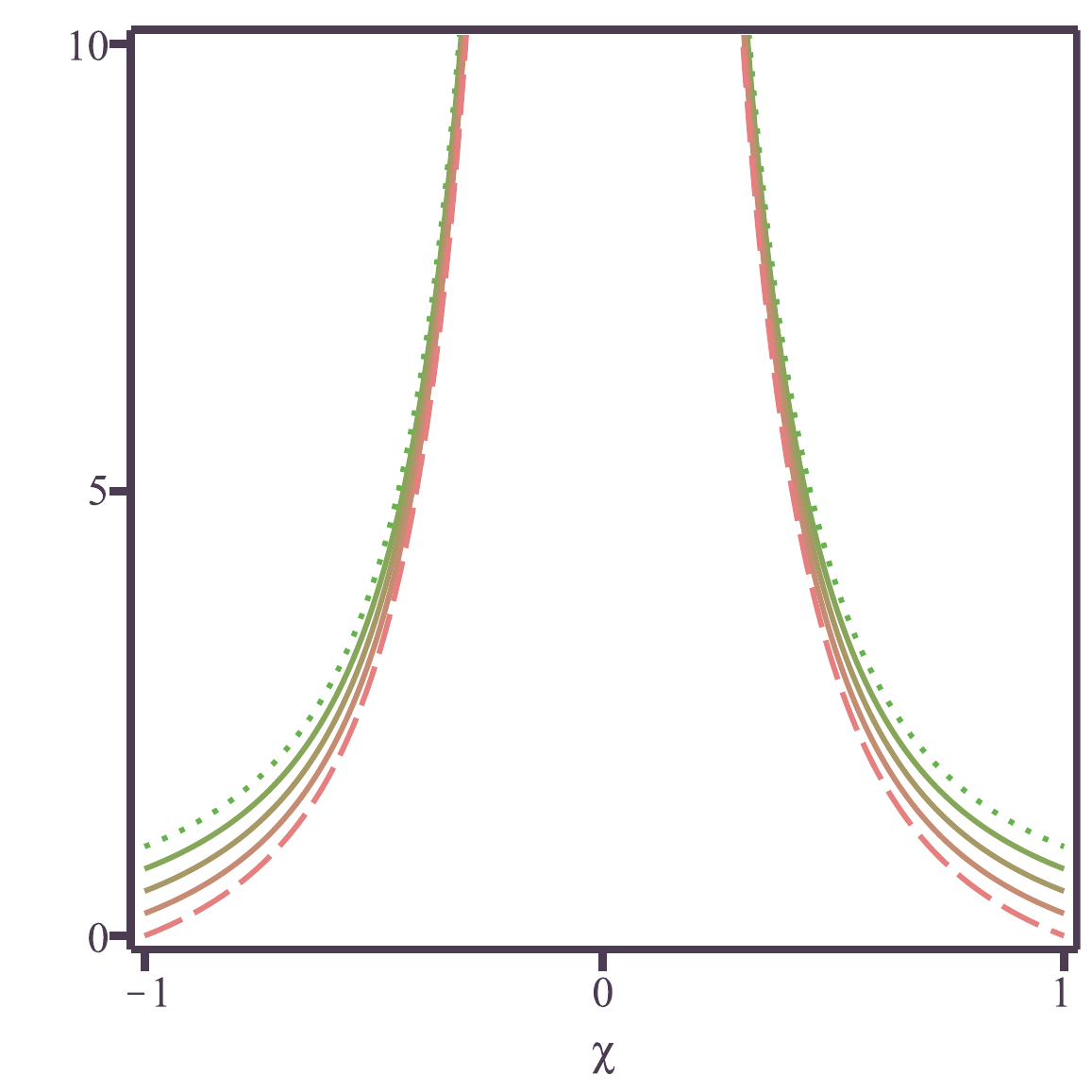}
	\includegraphics[width=6.2cm]{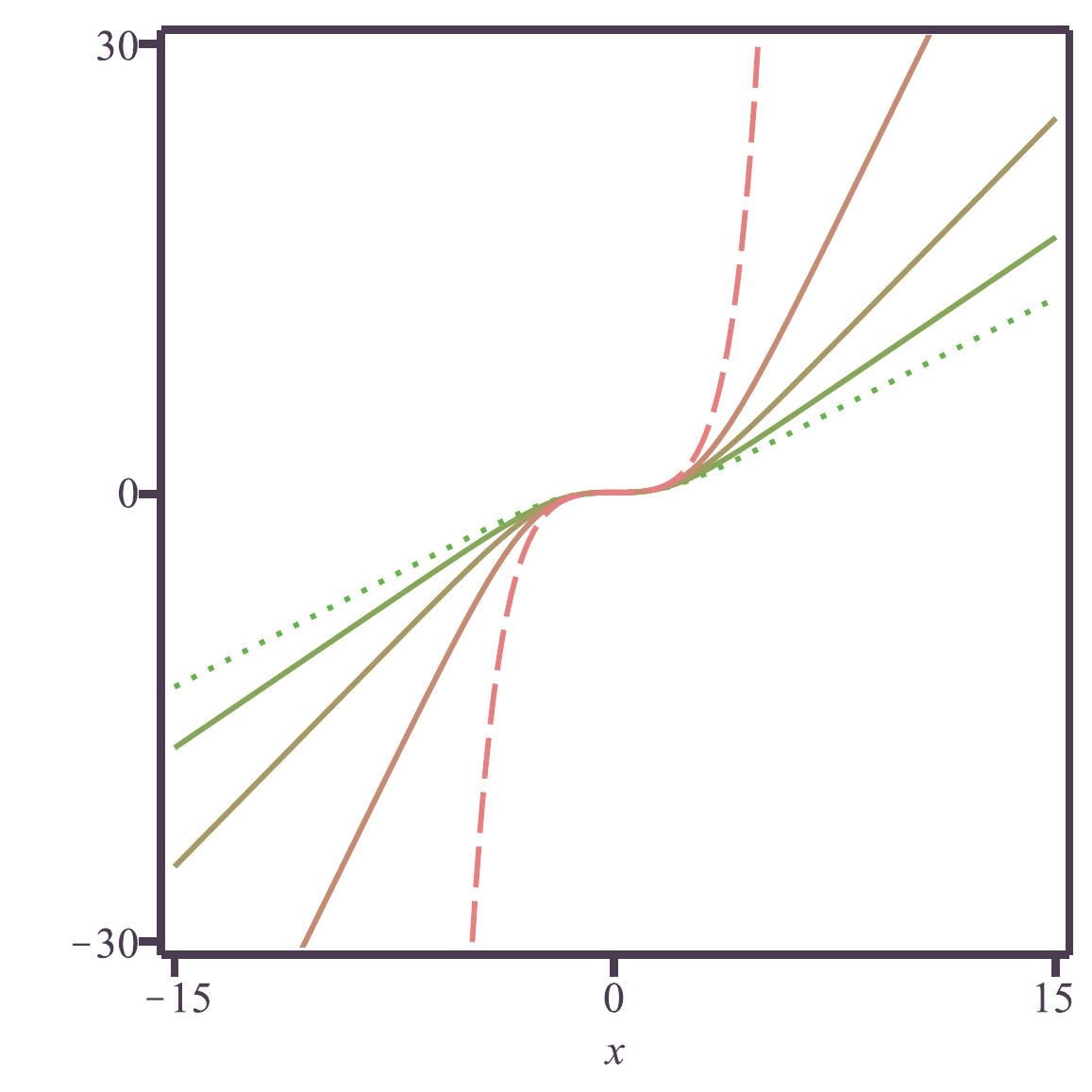}\\
	\includegraphics[width=6.2cm]{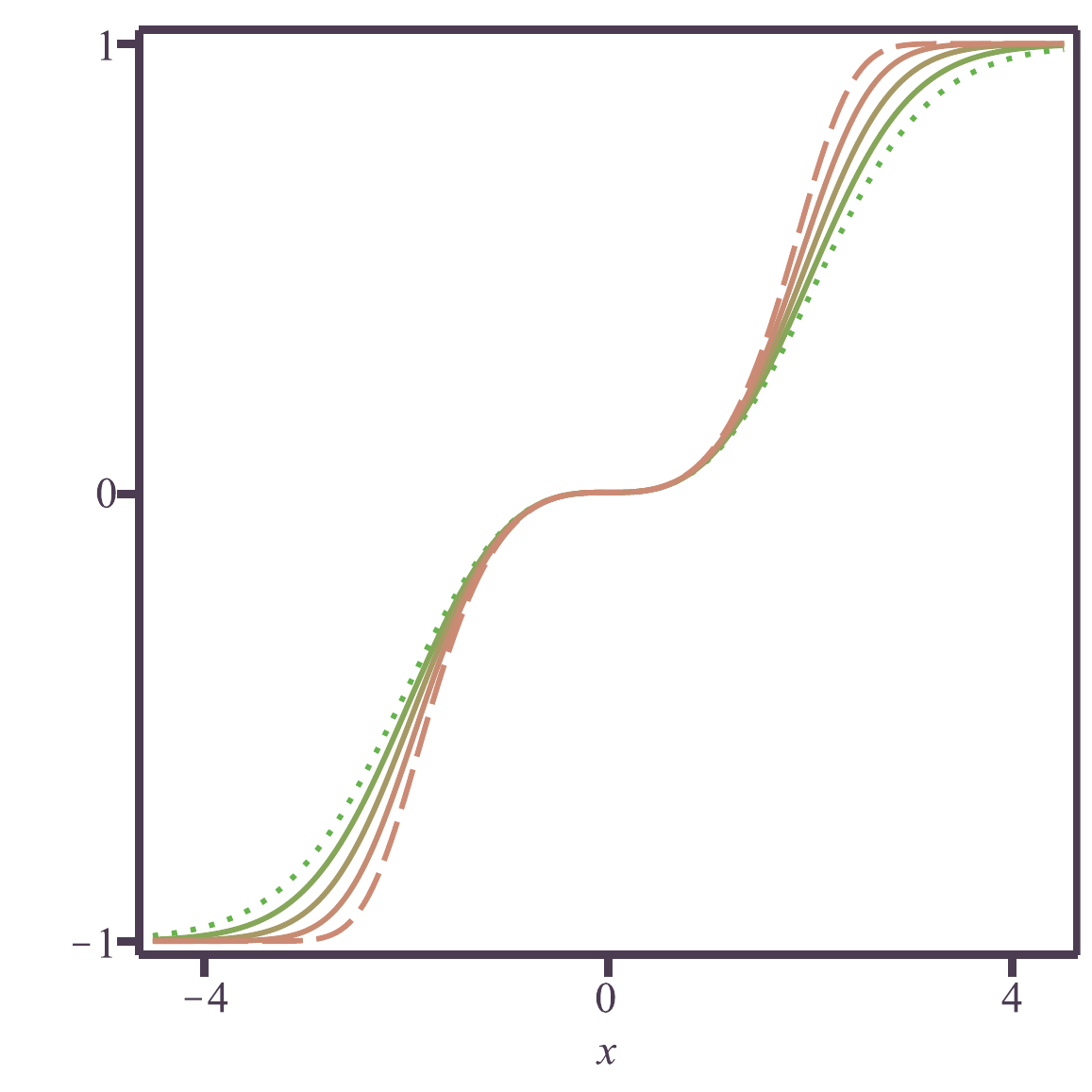}
	\includegraphics[width=6.2cm]{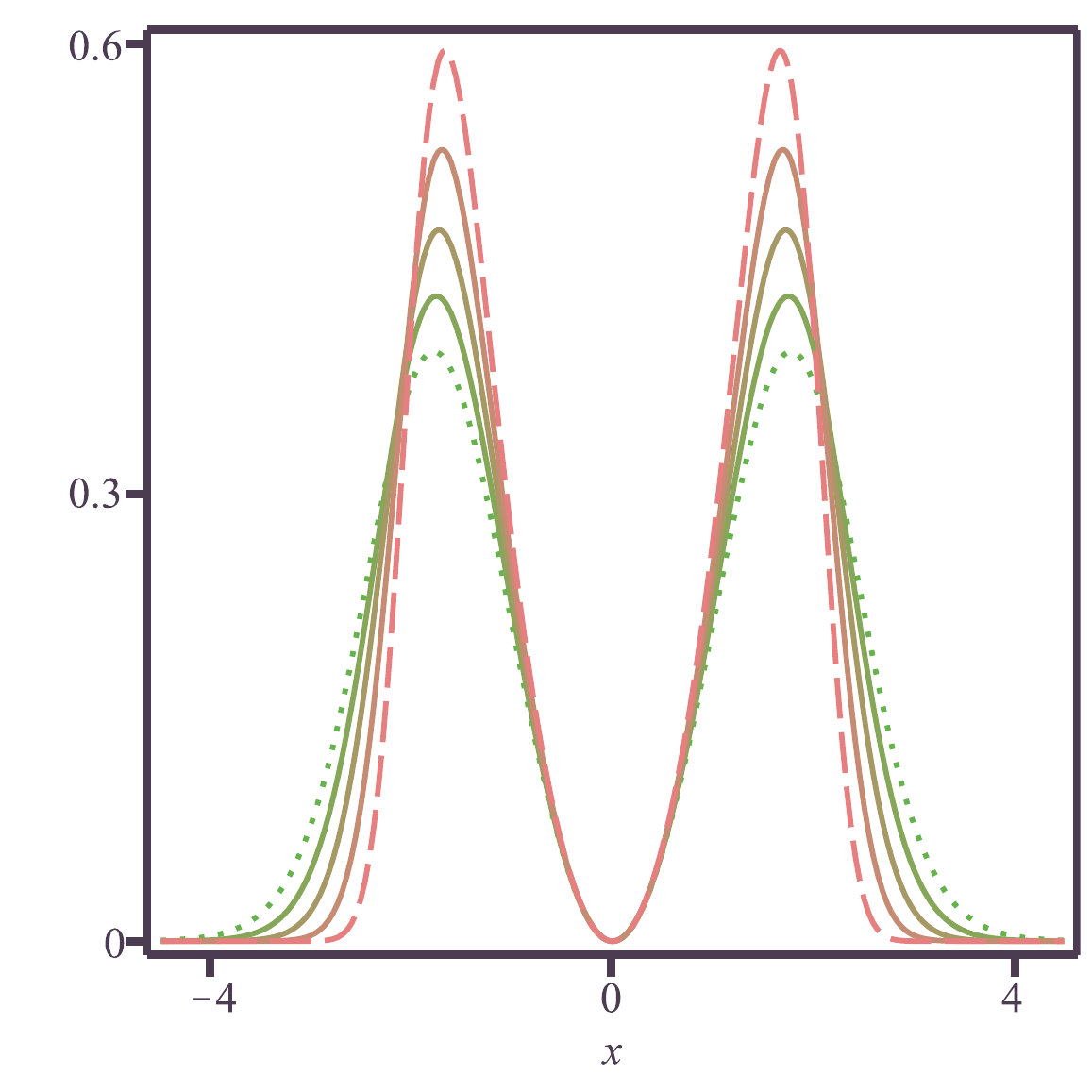}
	\caption{The function $f(\chi)$ (top left), the coordinate $\xi$ (top right), the solution $\phi$ (bottom left) and the energy density $\rho_1$ (bottom right) associated to the model described by Eqs.~\eqref{f2} and \eqref{gchi4} for $\alpha=1/2$ and for 
 $\beta=0, 0.25, 0.5, 0.75,$ and $\beta=1$. The dotted lines represent the case $\beta=0$ and the dashed ones stand for $\beta=1$.}
	\label{secondmodel}
\end{figure}

Let us now consider the opposite scenario, in which the intensity of the falloff is increased smoothly. To do this, we take
\be\label{f2}
f(\chi) = \frac{|1-\beta^2\chi^2|}{\chi^2},
\ee
with $0\leq\beta\leq 1$. The case $\beta=0$ recovers the model investigated in Ref.~\cite{multikink}, in which a double kink arise. To feed this function, we take $\chi$ driven by 
\be\label{gchi4}
g(\chi)=\alpha\chi-\frac\alpha3\chi^3.
\ee
For this choice, Eq.~\eqref{fochig} supports the solution
\be\label{solchi4}
\chi(x) = \tanh(\alpha x),
\ee
such that the energy density $\rho_2$ in Eq.~\eqref{rho12} reads
\be\label{rho2chi4}
\rho_2(x) = \alpha^2\sech^4(\alpha x).
\ee
Notice that $\chi(x)$ is now limited to the range $(-1,1)$. In the boundary values of $\chi(x)$, we have $f(\pm1) = 1-\beta^2$. So, we have $f(\chi)>1-\beta^2$ for $\chi\in(-1,1)$. Contrary to the previous case, higher values of $\beta$ makes the function becomes smaller, such that $f(\pm1) = 0$ for $\beta=1$. We display the results in Fig. \ref{secondmodel} for $\alpha=1/2$ and for $\beta=0, 0.25, 0.5, 0.75,$ and $\beta=1$. The behavior of $f(\chi)$ can be seen in the top left panel of Fig.~\ref{secondmodel}, where one shows that $f(0)$ is divergent, which makes the solution to develop the plateau near the origin, and the energy density $\rho_1$ the splitting behavior, shown in the bottom panels of Fig. \ref{secondmodel}, respectively.

We then use Eq.~\eqref{xidef} to find
\be\label{xi2beta}
\xi = \frac{1}{1-\beta^2}\left(x-\frac{\textrm{arctanh}\,(\beta\,\tanh(\alpha x))}{\alpha\beta}\right) +\xi_0
\ee
for $0<\beta<1$. The extreme values of $\beta$ lead to special forms of $\xi$. We have
\be
\xi = x-\frac{1}{\alpha}\tanh(\alpha x) + \xi_0
\ee
for $\beta=0$ and
\be\label{xidouble}
\xi = \frac{\sinh(2\alpha x)}{4\alpha} - x + \xi_0
\ee
for $\beta=1$. In the top right panel of Fig.~\ref{secondmodel}, we display the $\xi(x)$ coordinate. Also, since $\phi(x)$ is given by Eq.~\eqref{tanhxi}, we see that, for $0\leq\beta<1$, at very large values of $x$, the $x$ term dominates the expression \eqref{xi2beta} and the solution has exponential tails. For $\beta=1$, it gets double exponential tails, as one can see from Eq.~\eqref{xidouble}. In the bottom left panel of Fig.~\ref{secondmodel}, we show the solution $\phi(x)$. We notice that it engenders the same double kink profile of the model investigated in Ref.~\cite{multikink}, but the tail goes to the boundary values faster as we increase $\beta$. From the left equation of \eqref{rho12}, we get the energy density. It is 
\be
\rho_1(x) = \left(\coth^2(\alpha x)-\beta^2\right)\sech^4(\xi).
\ee
In the bottom right panel of Fig.~\ref{secondmodel}, we depict the above expression for $\rho_1(x)$. It engenders a double bell shape profile, which is directly connected with the double kink behavior of the solution $\phi(x)$ depicted in the bottom left panel of Fig. \ref{secondmodel}. This is an interesting splitting behavior, which may be further explored under collision.

As we can see, the model defined by the functions \eqref{f2} and \eqref{gchi4} leads to a falloff faster than the usual exponential one. 

\subsection{Solutions with compact behavior}

In the next models, we show that the exponential tail can be compactified. To do so, we take the function $g(\chi)$ in Eq.~\eqref{gchi4}, whose associated solution is \eqref{solchi4} with energy density \eqref{rho2chi4}. The modifications shall be then induced by $f(\chi)$.

\subsubsection{The case of half-compact behavior}

First, we investigate how to go from kinks to half-compact structures. We use $g(\chi)$ as in Eq. \eqref{gchi4} and take the coordinate $\xi$ to be driven by the following second-order polynomial function 
\be\label{f3}
f(\chi) = \frac{1+\beta^2\chi^2}{1+\beta^2},
\ee
where $\beta$ is a non-negative parameter. The decoupled case ($f=1$) is recovered for $\beta=0$. This function presents the feature of being minimum at the center of $\chi$, which is $x=0$. At this point, we have $f=1/(1+\beta^2)$, which tends to zero for $\beta\to\infty$. 
\begin{figure}[t!]
	\includegraphics[width=6.2cm]{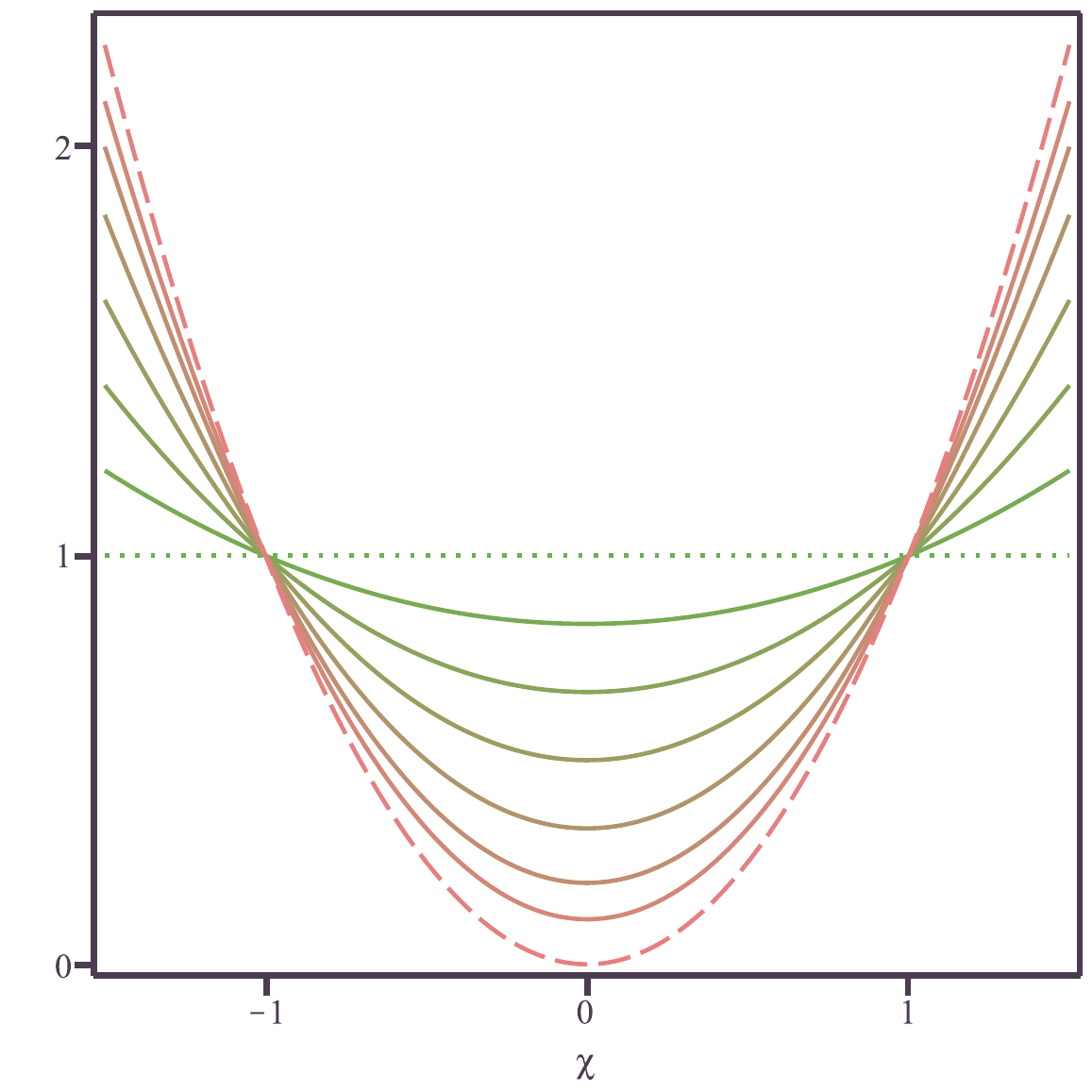}
	\includegraphics[width=6.2cm]{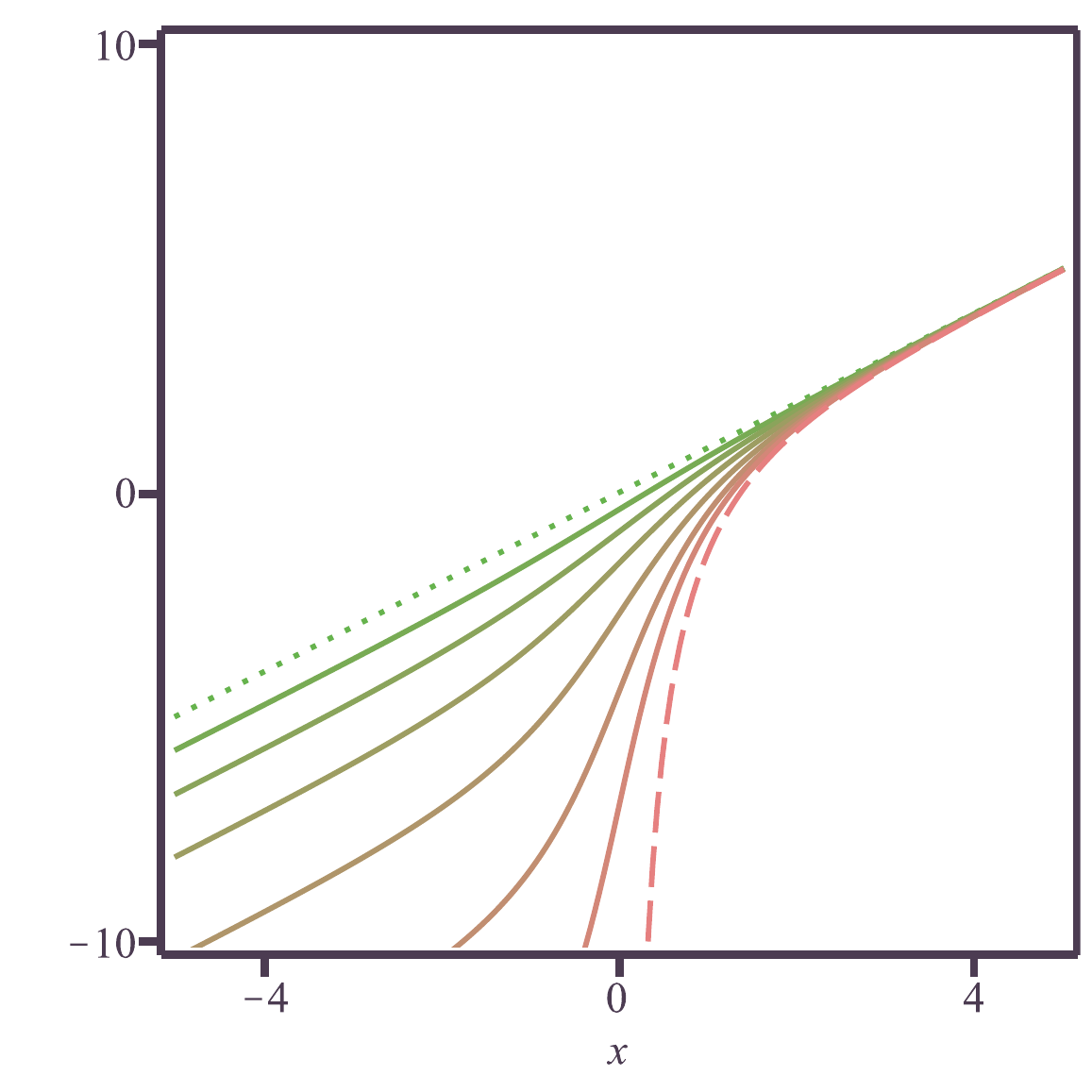}\\
	\includegraphics[width=6.2cm]{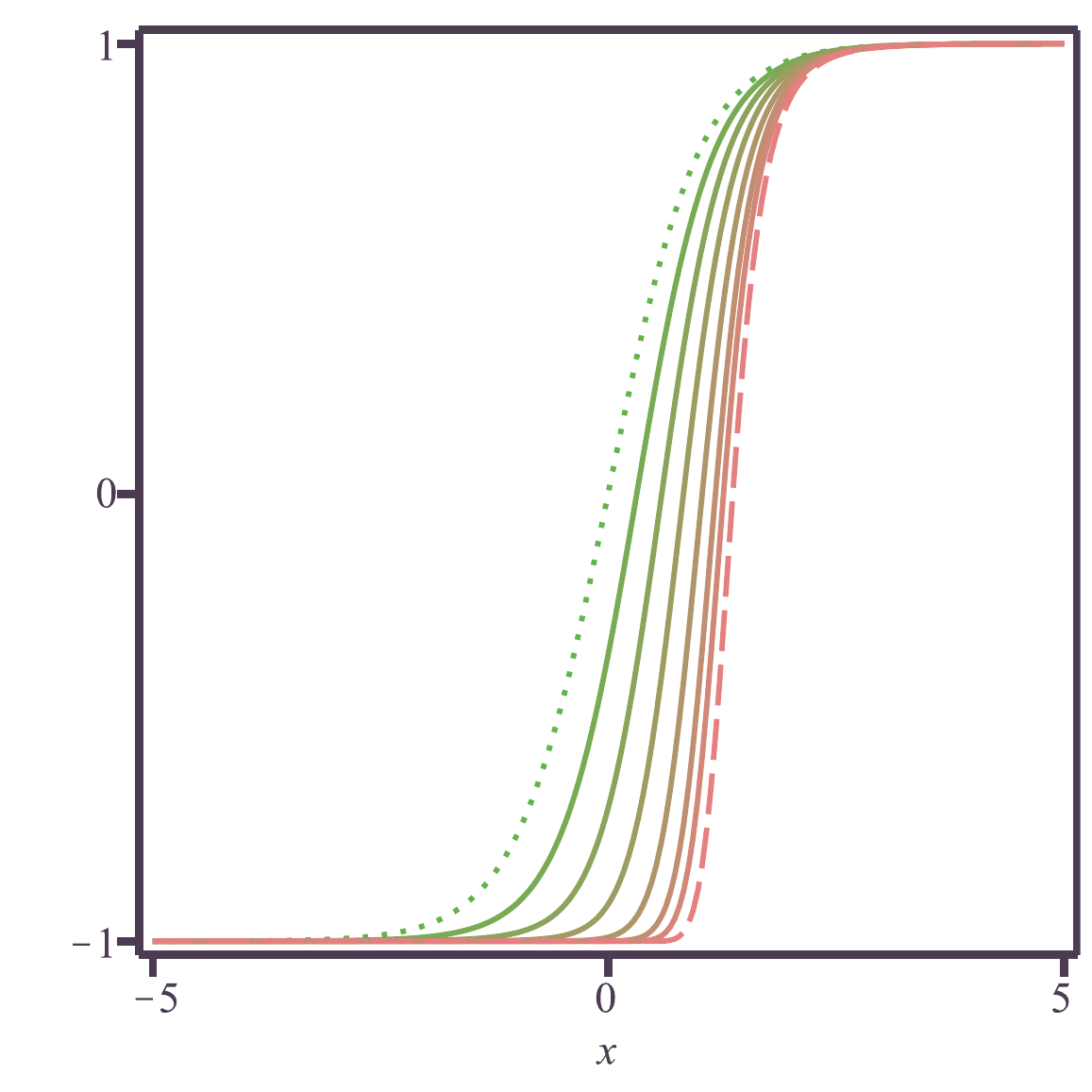}
	\includegraphics[width=6.2cm]{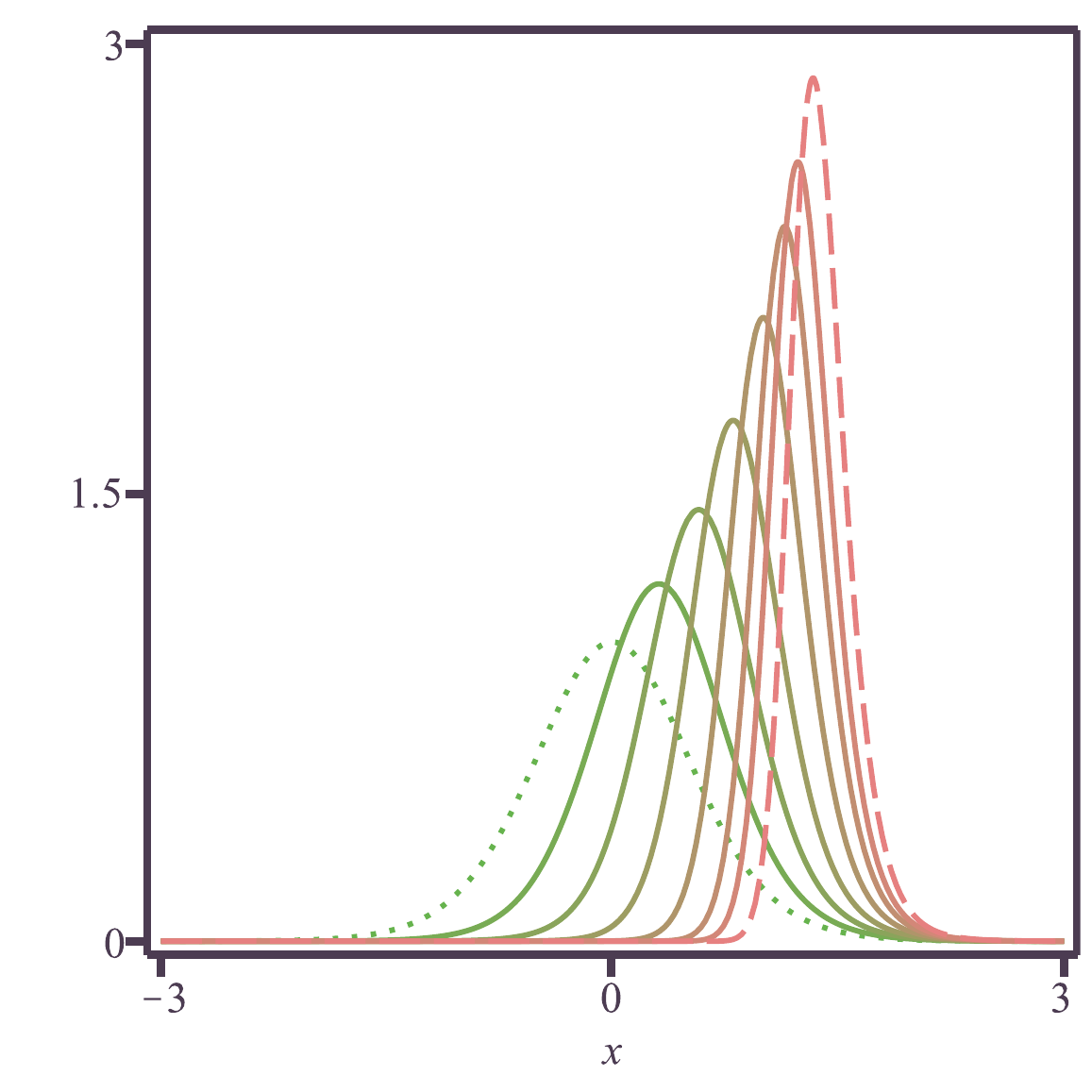}
	\caption{The function $f(\chi)$ (top left), the coordinate $\xi$ (top right), the solution $\phi$ (bottom left) and the energy density $\rho_1$ (bottom right) associated to the model described by Eqs.~\eqref{f3} and \eqref{gchi4} for $\alpha=1/2$, $c=2$ and $\beta=0, 0.2, 0.5, 1, 2, 4, 8,$ and $\beta\to\infty$. The dotted lines represent the case $\beta=0$ and the dashed ones stand for $\beta\to\infty$.}
\label{thirdmodel}
\end{figure}
In Fig. \ref{thirdmodel} we depict the results for $\beta=0, 0.2, 0.5, 1, 2, 4, 8$ and $\beta\to \infty$. In the top left panel of Fig.~\ref{thirdmodel}, we display the above function $f(\chi)$. 

From Eq.~\eqref{xidef}, we get
\be
\xi = x+ \frac{\beta\arctan(\beta\tanh(\alpha x))}{\alpha}-\frac{\beta\arctan(\beta)}{\alpha} + c,
\ee
where $c$ is a constant of integration. For large values of $x$, one sees from the above expression that the term linear in $x$ drives the coordinate $\xi$. So it behaves standardly. On the other hand, one must be careful when analyzing the point $x=0$, which leads to $\xi(0) = -\beta\arctan(\beta)/\alpha +c$. Here one finds that, as $\beta$ gets larger and larger, $-\xi(0)$ gets larger and larger. This means that $\beta\to\infty$, for which $f=\chi^2$, brings the tail at $x\to-\infty$ to $x=0$. In this limit, which brings the half-compact (hc) behavior, one can show that
\be
\xi_{hc} = \begin{cases}
\displaystyle x+\frac{1- \coth(\alpha x)}{\alpha} + c,\,\,&x> 0\\
-\infty, \,\, & x\leq0,
\end{cases},
\ee
To better illustrate the behavior of the $\xi$ coordinate, we depict it in the top right panel of Fig.~\ref{thirdmodel}. Moreover, to get the solution, which is depicted in the bottom left panel of Fig.~\ref{thirdmodel}, we must use the above coordinate in Eq.~\eqref{tanhxi}. Notice that, as $\beta$ gets larger and larger, the left tail of the solution gets a faster and faster falloff. In the limit $\beta\to\infty$, the solution attains the half compact profile
\be\label{phihc}
\phi_{hc}(x) = \begin{cases}
\displaystyle \tanh\left(x+\frac{1- \coth(\alpha x)}{\alpha} + c\right),\,\,&x> 0\\
-1, \,\, & x\leq0.
\end{cases}
\ee
The energy density associated to the solution for general $\beta$ can be calculated from Eq.~\eqref{rho12}, which reads
\be
\rho_1(x) = \frac{\left(1+\beta^2\right)\sech^4(\xi)}{1+\beta^2\tanh^2(\alpha x)}.
\ee
The energy density corresponding to the half-compact kink \eqref{phihc} is, then
\be
\rho^{hc}_1(x) = \begin{cases}
\displaystyle \coth^2(\alpha x)\,\sech^4\left(x+\frac{1- \coth(\alpha x)}{\alpha} + c\right),\,\,&x> 0\\
0, \,\, & x\leq0.
\end{cases}
\ee
In the bottom right panel of Fig.~\ref{thirdmodel}, we display the above energy density for several values of $\beta$. Notice that, as $\beta$ increases it gets taller and narrower, since the left tail shrinks toward zero.

\subsubsection{The case of compact behavior}

\begin{figure}[t!]
	\includegraphics[width=6.2cm]{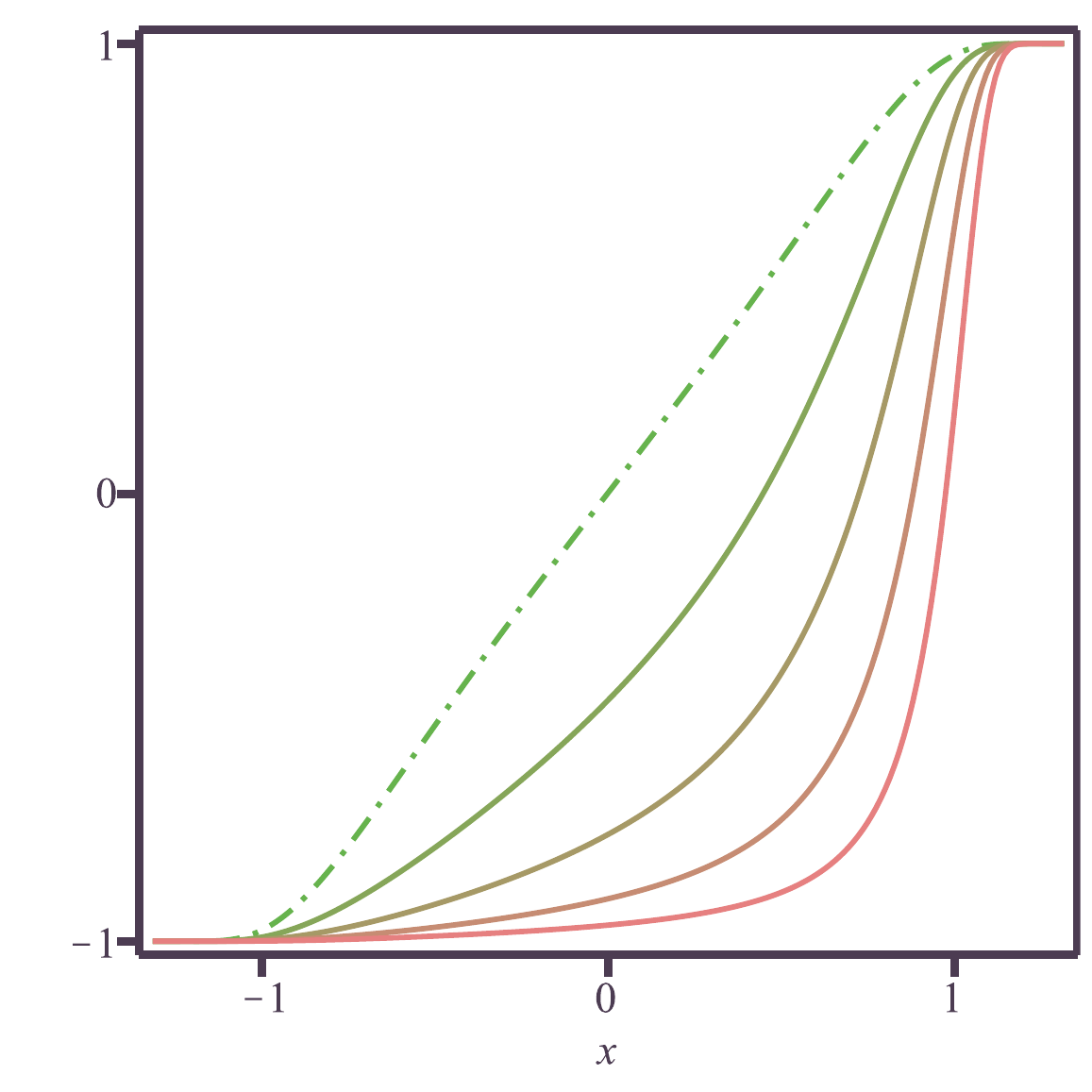}
	\includegraphics[width=6.2cm]{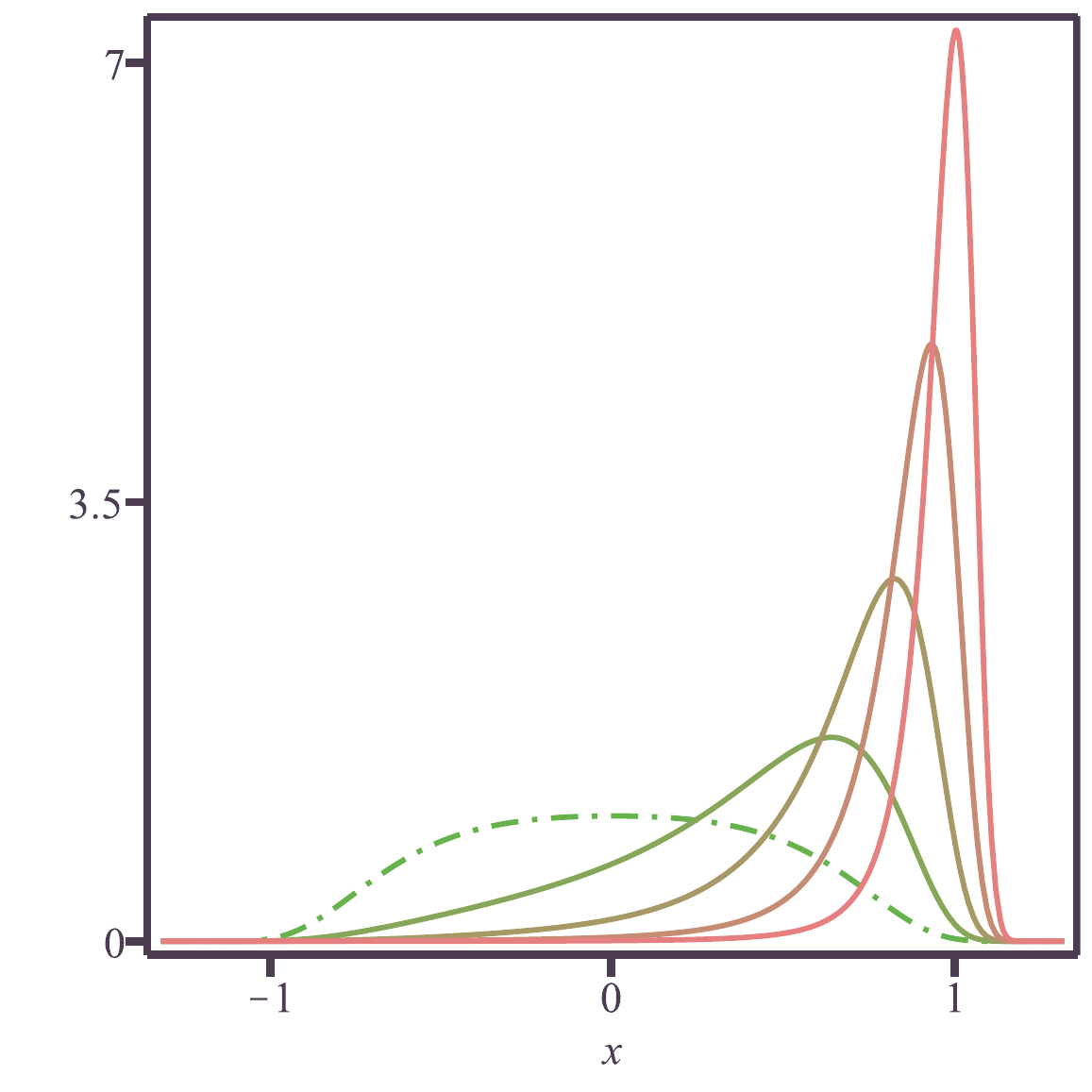}\\
	\includegraphics[width=6.2cm]{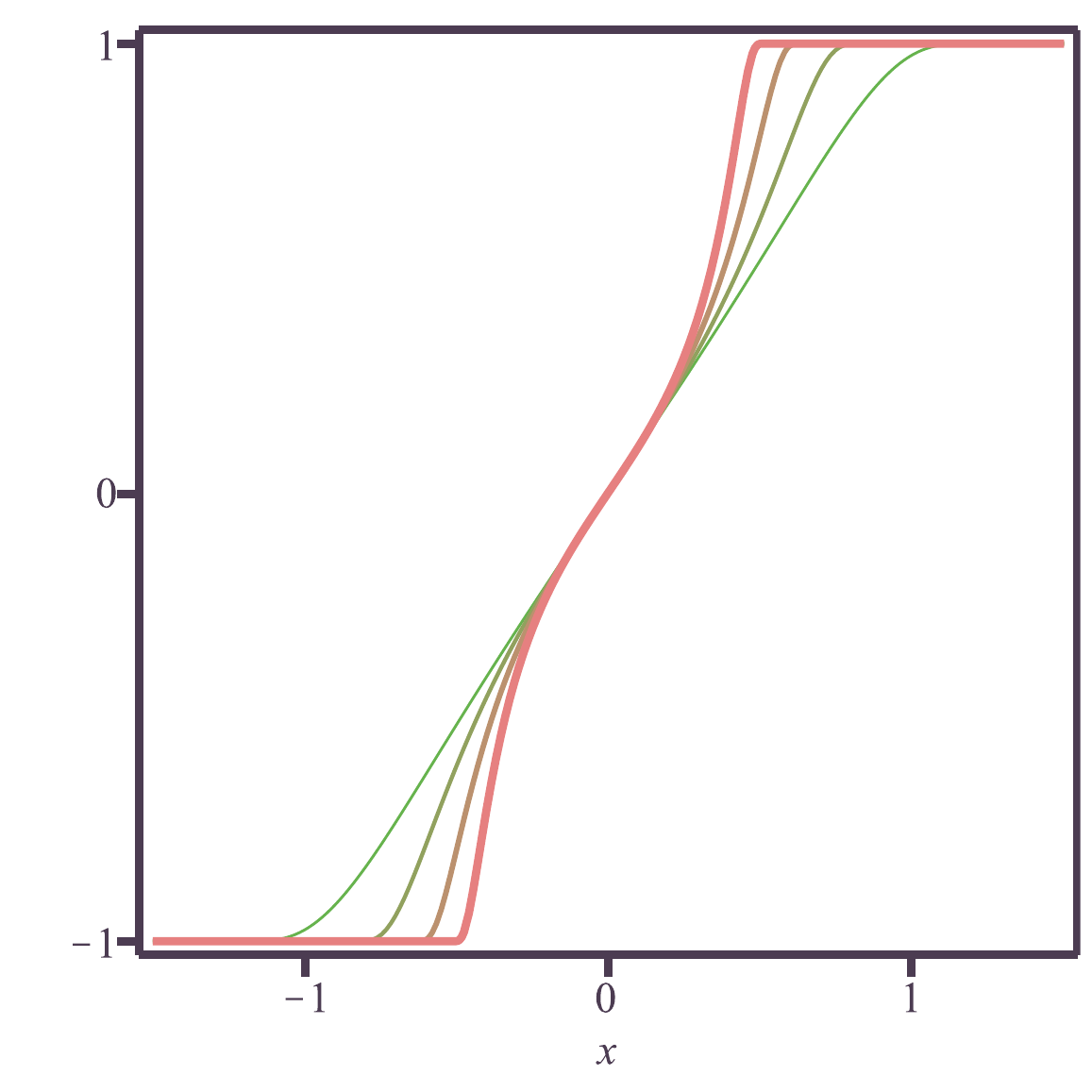}
	\includegraphics[width=6.2cm]{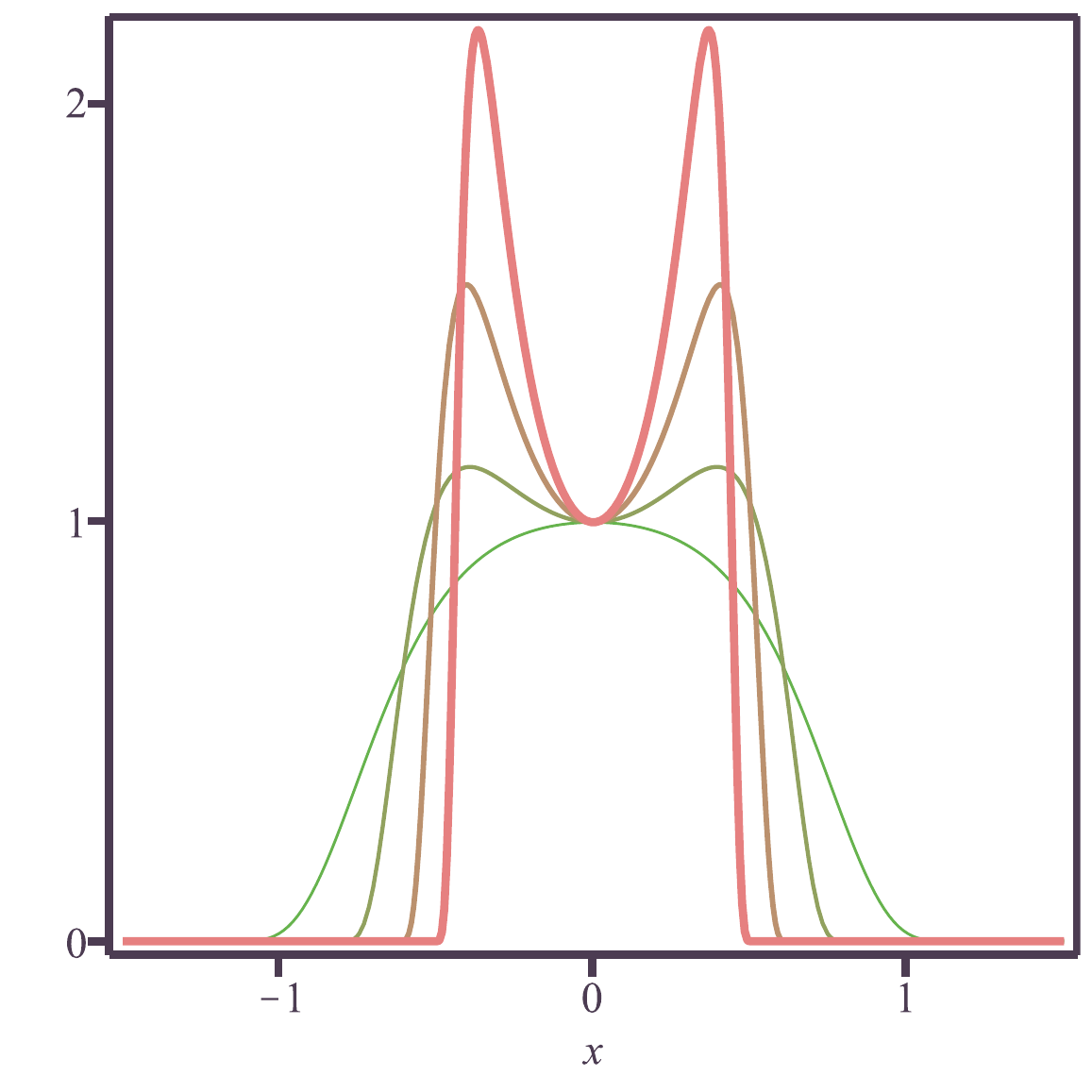}
	\caption{The solution $\phi$ (left) and the energy density $\rho_1$ (right) associated to the model described by Eqs.~\eqref{f4} and \eqref{gchi4} for $\alpha=2$ and $c=0,0.5,1,1.5,2$ (top), and $c=0$ and $\alpha=0.5,0.75,1,1.25$ (bottom). In the top panels, the dash-dotted lines represent the case $c=0$. In the bottom panels, the thickness of the lines increases with $\alpha$.}
	\label{fourthmodel}
\end{figure}

In the previous model, we saw that, due to Eq.~\eqref{xidef}, the presence of a zero in $f(\chi)$ at $\chi=0$, led to a half compact kink. In order to compactify both tails of the solution, we present a model which supports two symmetric zeroes in $f(\chi)$. It arises for $g(\chi)$ as in Eq.~\eqref{gchi4} and the new choice
\be\label{f4}
f(\chi) = (1-3\chi^2)^2.
\ee
The minima of this function are located at $\chi=\pm\sqrt{3}/3$, which are also its zeroes. From Eq.~\eqref{xidef}, one gets
\be
\xi = \begin{cases}
\displaystyle\frac{x}{4} - \frac{3\tanh(\alpha x)}{4\alpha (3\tanh^2(\alpha x) -1)} -c,\,\,&|x|\leq x_c\\
\textrm{sgn}(x) \,\infty, \,\, & |x|>x_c
\end{cases},
\ee
where $x_c=\textrm{arctanh}(1/\sqrt{3})/\alpha$. By using this expression in Eq.~\eqref{tanhxi}, one gets the compact solution
\be\label{phicomp}
\phi(x) = \begin{cases}
\displaystyle\tanh\left(\frac{x}{4} - \frac{3\tanh(\alpha x)}{4\alpha (3\tanh^2(\alpha x) -1)} -c\right),\,\,&|x|\leq x_c\\
\textrm{sgn}(x), \,\, & |x|>x_c
\end{cases},
\ee
The energy density in Eq.~\eqref{rho12} takes the form
\be\label{rhocomp}
\rho_1(x) = \begin{cases}
\displaystyle\left(2-3\,\sech^2(\alpha x)\right)^{-2}\,\sech^4\left(\frac{x}{4} - \frac{3\tanh(\alpha x)}{4\alpha (3\tanh^2(\alpha x) -1)} -c\right),\,\,&|x|\leq x_c\\
0, \,\, & |x|>x_c
\end{cases}.
\ee
 In Fig.~\ref{fourthmodel}, we depict the solution in Eq.~\eqref{phicomp} and its energy density in \eqref{rhocomp} for $\alpha=2$ and $c=0,0.5,1,1.5,2$ in the top panels, and for $c=0$ and $\alpha=0.5,0.75,1,1.25$, in the bottom panels. We notice that the parameter $c$ introduces an asymmetry in the system, shifting the peak of the energy density. Also, for $c=0$, the energy density may split, with two maxima arising and becoming more apparent as $\alpha$ gets larger. This is another interesting splitting behavior, which may be further explored under collision.

We remark that the expressions in Eqs.~\eqref{f4}--\eqref{rhocomp} only show the case in which compact solutions arise. However, one may introduce a parameter that smoothly connects the decoupled case ($f=1$) to the one in Eq.~\eqref{f4}. We did not investigate this here because the study is similar to the ones described in the previous models.

\section{Ending comments}

In this work, we have investigated how the presence of a field $\chi$ that drives the function $f(\chi)$ in \eqref{lagr} can modify the asymptotic behavior of $\phi$. It is known from Refs.~\cite{multikink,multikink2} that it can induce a distinct geometric coordinate in kink solutions, simulating the effects of geometric constrictions \cite{jubert}. Based on the energy minimization, we have used first order equations in which the $\chi$ field is independent from $\phi$ to show that, to modify the tails of a kinklike solution, one must take $\chi$ and $f(\chi)$ that leads to a coordinate $\xi$ in Eq.~\eqref{xidef} ranging from $-\infty$ to $+\infty$.

In the first model, we have considered that $\chi$ is the vacuumless solution \eqref{solchi1} that feeds the function \eqref{f1}. In this case, the parameter $\beta$ smoothly controls the transition between the standard solutions ($\beta=0$), with exponential tails, to the long range ones ($\beta\to\infty$), in which the falloff is given in terms of a power-law expression. Next, we have taken the well-known hyperbolic tangent for $\chi$; see Eq.~\eqref{solchi4}. In the second model, described by the function \eqref{f2}, we have shown how to increase the intensity of the falloff of the solution, going from the standard to the double exponential profile. The third model, described by \eqref{f3} introduce a novel procedure to compactify one of the tails of a kink. In this case, a parameter controls the minimum of $f(\chi)$ making it becoming a zero for $\beta\to\infty$. In the fourth model \eqref{f4}, we present a function that compactifies both tails of the kink solution. We highlight that this is a novel mechanism to get compact structures, very different from the ones previously investigated in Refs.~\cite{comp,kinktocomp,hybrid}. In the four models, we considered for $\phi$ the same contribution $\phi-\phi^3/3$ to $W(\phi,\chi)$, as introduced in Eq. \eqref{W}, but we have used different $f(\chi)$ and $g(\chi)$. In particular, in the first model we considered $g(\chi)=\alpha\,\arctan(\sinh(\chi))$ and $f(\chi)$ given in \eqref{f1}, and in the three last models we considered the same $g(\chi)=\alpha\chi-\alpha\chi^3/3$, but different $f(\chi)$, as in \eqref{f2}, \eqref{f3} and \eqref{f4}, to show the importance of this function in the construction of geometrically constrained configurations. These results add to the previous ones, introduced in Refs. \cite{multikink,multikink2}, to strengthen the effects imposed by $f(\chi)$ in the Lagrangian density \eqref{lagr} in the search for localized structures.

A direct perspective concerns the use of different functions $f(\chi)$ and distinct potentials $V(\phi,\chi)$, to find  
other solvable models. We can, for instance, change $\phi$ to a couple of fields $(\phi_1,\phi_2)$ to see how the $\chi$ field can be used to constrain the pair $(\phi_1,\phi_2)$. There are several other possibilities, in particular, we can think of using the above new solutions in collisions, to see how they respond to the scattering process. 
This investigation can be done following the lines of Refs. \cite{gani,sca1,S1,S2,S3} and references therein. In particular, in the energy densities of the second and fourth models, depicted in Figs. \ref{secondmodel} and \ref{fourthmodel}, respectively, we identified the splitting behavior, which may induce fragmentation of the localized structure. This means that such structures may give rise to unusual behavior in the scattering process, a possibility which is presently under consideration. We can also consider more general models, including complex scalar and gauge fields, to find other localized structures and to investigate how the geometric constriction that appears in one spatial dimension can be extended to higher dimensions.

\vspace{1cm}
{\noindent\textbf{Data availability statement}: This manuscript has no associated data.}

\acknowledgements{This work is supported by the Brazilian agency Conselho Nacional de Desenvolvimento Cient\'ifico e Tecnol\'ogico (CNPq), grants Nos. 303469/2019-6 (DB), 306151/2022-7 (MAM) and 310994/2021-7 (RM). It is also supported by Paraiba State Research Foundation (FAPESQ-PB), grants Nos. 0003/2019 (RM) and 0015/2019 (DB and MAM).}


\end{document}